\documentclass[aps,prl,reprint,showpacs,amsmath,amssymb]{revtex4-1}

\usepackage{graphicx}
\usepackage{bm}

\begin{document}
\title{Experimental scaling law for the sub-critical transition to turbulence in plane Poiseuille flow}
\author{Gr\'egoire Lemoult}
\email[]{gregoire.lemoult@espci.fr}
\author{Jean-Luc Aider}
\email[]{jean-luc.aider@espci.fr}
\author{Jos\'e Eduardo Wesfreid}
\email[]{wesfreid@pmmh.espci.fr}

\affiliation{PMMH (UMR7636 CNRS - ESPCI - UPMC Univ. Paris 6 - UPD Univ. Paris 7), \'Ecole Sup\'erieure de Physique et de Chimie Industrielles, 10 rue Vauquelin, 75005 Paris, France}

\date{\today}

\begin{abstract}
We present  an experimental study of transition to turbulence in a plane Poiseuille flow. Using a well-controlled perturbation, we analyse the flow using extensive Particule Image Velocimetry and flow visualisation (using Laser Induced Fluorescence) measurements and use the deformation of the mean velocity profile as a criterion to characterize the state of the flow. From a large parametric study, four different states are defined depending on the values of the Reynolds number and the amplitude of the perturbation. We discuss the role of coherent structures, like hairpin vortices, in the transition. We find that  the minimal amplitude of the perturbation triggering transition scales like $Re^{-1}$.
\end{abstract}


\maketitle

For more than a century \cite{reynolds1883experimental}, the transition to turbulence in shear flows has been a prolific domain of study. Despite many theoretical investigations \cite{grossmann2000onset}, it has not been possible to predict correctly this transitional process even for flows in simple geometries such as circular or plane Poiseuille flow, or plane Couette flow. Nowadays, the transition process in pipe and channel flows remains one of the most fundamental and practical problems still unsolved in fluid dynamics.

Linear stability theory has been applied to plane Poiseuille flow, linearizing the Navier-Stokes equation near the stable parabolic profile. The smallest unstable or critical Reynolds number obtained was $Re_c=5772$ ($Re=u_{cl}h/\nu$ with $u_{cl}$ the laminar center line velocity, $h$ the half channel height and $\nu$ the kinematic viscosity of the fluid) \cite{lin1946, Orzag}. This result is in contradiction with the experimental work of Carlson \textit{et al.} who found $1000<Re_c<2000$ \cite{carlson1982flow}.

Other approaches, related to the non-normal character of the Navier Stokes operator linearized around the stable laminar flow solution, can support important transient growth of finite amplitude disturbances, related to  streamwise and quasi-streamwise alignment \cite{ trefethen1993hydrodynamic, chapman2002subcritical}. 

Waleffe, from an 3D non-linear  modal reduction of the Navier-Stokes equations, adopted the idea of a self sustained process as origin of the transition \cite{waleffe1997self}. After destabilization of these streaks, streamwise modes appear and regenerate the vortices through a non-linear interaction, as experimentally observed by Duriez \textit{et al.} \cite{duriez}.

Transition to turbulence of sheared flows, is characterized by a double threshold, in the initial disturbance and in the Reynolds number, where the larger the Reynolds number, the smaller  the necessary perturbation. This behavior, when the flow becomes unstable with respect to finite amplitude disturbances, is described with a power law for the minimal amplitude of the disturbance triggering the transition: $$ \epsilon = O(Re^\gamma) $$

After the first studies of Trefethen \cite{trefethen1993hydrodynamic} on the critical exponent $\gamma$, Chapman studied the transient growth of sub-critical transition process in the plane Poiseuille flow \cite{chapman2002subcritical}. He found $\gamma=-3/2$ for initial streamwise vortices and $\gamma=-5/4$ for initial oblique vortices.

On the other hand, the results of the nonlinear study of Waleffe and Wang show that $\gamma=-1$ in shear flows \cite{waleffe2005transition}. Numerical experiments of Kreiss \textit{et al.} give $-21/4 \leq \gamma \leq -1$ \cite{kreiss1994bounds}.

Critical exponents were measured experimentally in plane Couette flow by Dauchot and Daviaud \cite{dauchot1995finite}, in pipes by Mullin \textit{et al.} \cite{darbyshire1995transition,hof2003scaling}  
and in channel flow by Ben-Dov \textit{et al.} \cite{cohen_2007}.
The latter found, using a cross jet as disturbance, that the critical jet velocity triggering the appearance of hairpin vortices scales like $Re^{-3/2}$. 

\begin{figure*}[htb]
\includegraphics[width=\textwidth]{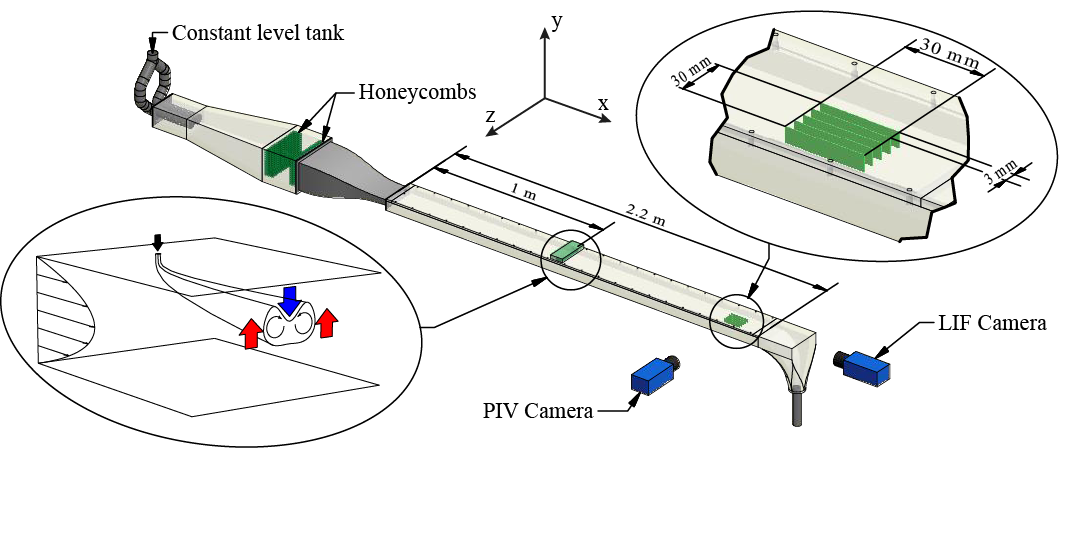}
\caption{\label{fig:SetUp} Schematic view of the dedicated water channel used in PMMH.  The developement section is $1$ m long and the test section is $1.2$ m long. Its cross section is $20\times150$ mm$^2$. Left-bottom, scheme of the perturbation generation by continuous injection of water through small holes in the upper plate. Arrows represents the lift-up effect generating high and low speed streaks. Top-right, scheme of the multi plane PIV technique used to compute the average velocity profile. We used 11 planes separated by $3$ mm.}
\end{figure*}

The present work studies experimentally the sub-critical transition of a plane Poiseuille flow perturbed by streamwise vortices induced by jets in cross flow. We will study the transition of the Poiseuille flow using the deformation of the mean velocity profile as a criterion instead of focusing on the apparition of hairpin vortices which is a step in the transition process.

The mean flow distortion was described in channels by Eliahou \textit{et al.} \cite{eliahou1998laminar}. It is indeed now well established, that the mean flow distortion is related to the presence of streamwise and quasi-streamwise elongated structures in the wall regions of boundary layers \cite{duriez2006base}, pipes and channels. These structures, alternating low and high momentum contributions to the flow, generate through a nonlinear coupling, a global modification of the flow. Recently, Barkley gave a nearly  complete model of the transition in pipe flow, introducing the modification of the mean velocity profile as one of the main ingredients of a system of coupled non-linear equations \cite{barkley2011simplifying}.

When comparing experimental data to models, it is difficult to define properly the amplitude of the perturbation \cite{citerapport} and also its critical value triggering the transition. Our choice of selecting the mean flow distortion to characterize the transition puts a special emphasis on a more rigorous definition of the onset of turbulence.

The experimental system is composed of a 3 m long plexiglass channel (Figure~\ref{fig:SetUp}). The test section's half height is $h=10$mm, its length $220 h$ and its width $15h$. The perturbation is generated $100h$ downstream from the inlet to ensure a fully developed Poiseuille flow for all $Re$. The $x,y$ and $z$ axis are respectively the streamwise,  normal to the walls and spanwise coordinates, with $y=0$ in the middle of the channel and $x=0$ where perturbations are injected. The design of the inlet section, together with the smooth connections between all parts of the channel, minimize the upstream perturbations leading to a laminar base flow until at least $Re=5500$ (The Reynolds number is estimated from volume flow measurements).

The flow is perturbed by \textit{continuous} injection of water through four circular holes, normal to the flow, drilled into the the upper wall with diameter $d = 0.2 h$ and spacing $\lambda = 3 h$. The structure of the flow induced by the jets may be complex and depends strongly on the amplitude of the perturbation \cite{karagozian2010transverse,ilak2010stability} defined as the velocity ratio $A=u_{jet} / u_{cl}$, where $u_{jet}$ is the mean jet velocity and $u_{cl}$ the unperturbed centerline velocity  .  For $0<A<2$, one can consider that each jet creates a pair of counter-rotating streamwise vortices (figure~\ref{fig:SetUp}) similar to those created by solid vortex generators in a flat-plate boundary layer \cite{duriez}. It has been shown that the destabilization of the streaks is a key step in a self-sustaining process between streaks and streamwise vortices in Poiseuille flow as well as in a flat-plate boundary layer  \cite{waleffe2001exact,duriez}. It is also a key step in the transition scenario proposed by Chapman\cite{chapman2002subcritical}.

\begin{figure}[htb]
\includegraphics[width=\columnwidth]{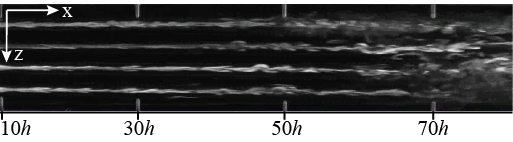}
\caption{\label{fig:Visu_Transition}LIF visualization of the transition in the $y=0$ plane. The flow goes from left to right.}
\end{figure}

\begin{figure*}[t]
\includegraphics[width=\textwidth]{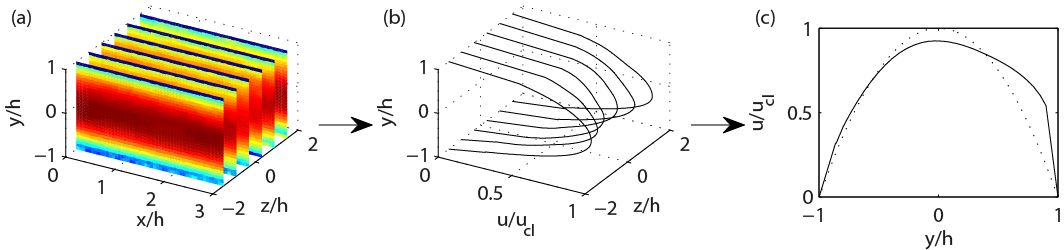}
\caption{\label{fig:Process}From left to right, decomposition of the  steps to compute the mean velocity profile. First we measure the time averaged (30 snapshots at $4Hz$) velocity field in 11 $x-y$ planes (only 6 presented here), then we average in the $x$-direction and finally, we average in the $z$-direction.}
\end{figure*}

We show on figure~\ref{fig:Visu_Transition} an exemple of visualization, obtained by Laser Induced Fluorescence (LIF), of the transition induced by the jets relatively far downstream from the injection. For $x<60h$, one can see that the streamwise vortices induced by the four jets are already unstable with clear streamwise modulation characteristic of hairpin vortices. This correspond to the mixed behaviour observed by Tasaka \textit{et al.} \cite{Tasaka2010}. The transition to turbulence occurs further downstream ($x>60h$) over the entire channel width. In the following, all measurements will be carried on in the region $75h<x<80h$ illustrated on figure \ref{fig:SetUp}.

The velocity field is studied using Particle Image Velocimetry (PIV) measurements (figure~\ref{fig:SetUp} and \ref{fig:Process}). The fluid is seeded with neutrally buoyant particles ($d_p\approx5 \mu  {\rm m}$). To take into account the spanwise modulation of the flow, we measure velocity fields in a volume ($\Delta x=3h$, $\Delta z=\lambda=3h$, $\Delta y=2h$), centred around $x=78h$. The measurement volume is divided in 11$x-z$ planes between $z=\pm \lambda/2$ around a jet. For each plane, 30 instantaneous snapshots are taken at $4$Hz giving 11 time-averaged velocity fields (figure \ref{fig:Process}.a). Then the PIV fields are space averaged over the streamwise direction (figure~\ref{fig:Process}.b). Finally, the 11 profiles are averaged along the spanwise direction giving the final mean velocity profile (figure~\ref{fig:Process}.c). Each mean profile presented in the following is the result of an average over $26000$ profiles. The total time to measure one mean profile is 3 minutes,  thus averaging intermittent effects.

\begin{figure}[htb]
\includegraphics[width=\columnwidth]{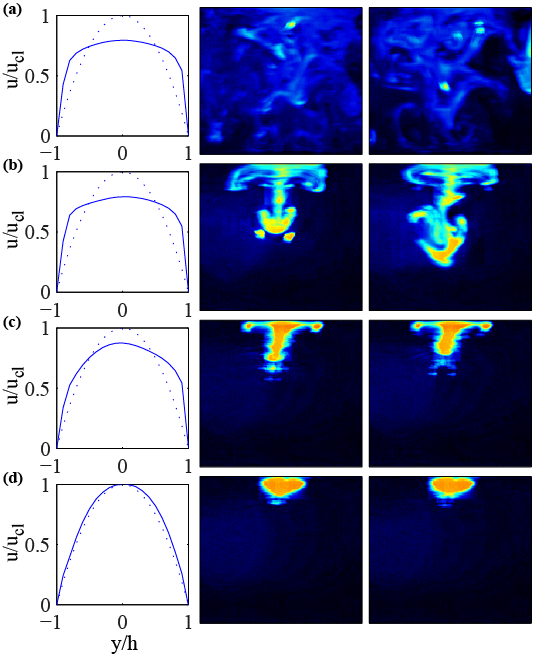}
\caption{First column represent the evolution of the mean velocity profiles, measured by PIV at $x=80h$, from which the parameter $\tilde{u}$  is defined and in following columns two snapshots taken using the  LIF technique,  at different times, in a $y-z$ plane located at $x=80h$. From top to bottom, rows correspond respectively to $A=$ 0.276, 0.248, 0.209 and 0.167, $Re=2830$ for all rows.}\label{fig:EvolutionVisu}
\end{figure}

\begin{figure}[t]
\includegraphics[width=\columnwidth]{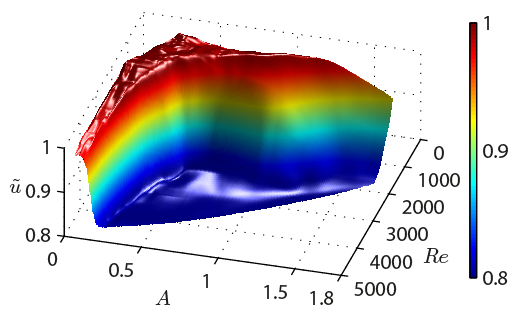}
\caption{3D plot of the surface representing the parameter $\tilde{u} = \max(u)/u_{cl}$ as a function of $Re$ and $A$ showing well separated domains of each state: laminar and turbulent.}
\label{fig:Surface3D}
\end{figure}

The transition from the laminar to the turbulent regime is associated with a deformation of the mean velocity profile from a parabolic to a plug profile (figure \ref{fig:EvolutionVisu}). The profile can then be used as a quantitative criterion to define the different states of the flow. We construct a state parameter $\tilde{u}$ as: $$ \tilde{u} = \frac{max(u)}{u_{cl}} $$ where $u$ is the perturbed mean velocity profile.

We plot on figure \ref{fig:Surface3D} the evolution of $\tilde{u}$ as a function of $Re$ ($500<Re<5300$) and $A$ ($0<A<2$). The state plane is sampled with 450 points. The sampling has been refined in the transition region. Two different domains or plateaux can be clearly identified: the laminar (red) and turbulent (blue) state, separated by a sharp cliff, giving a description of the global state of the flow. We now need to define a rigorous criterion to identify the transition.

For this purpose, we visualized the flow using LIF in the $x=80h$ plane. The structure of the flow is clearly modified when the amplitude of perturbation is increased. It is illustrated on figure \ref{fig:EvolutionVisu} for $Re=2830$. For a small perturbation (fourth row, $A = 0.167$), the flow is steady and laminar. Even if the fluorescent patches show traces of hairpin vortices, the mean velocity profile remains parabolic ($\tilde{u}=1$). As the amplitude of the perturbation is increased (third row, $A = 0.209$), hairpin vortices becomes larger but still smaller than the channel's half-height and slightly unsteady. In this case the mean velocity profile becomes asymmetric and $\tilde{u}$ decreases ($\tilde{u}=0.9$). For $A = 0.248$ (second row), hairpin vortices becomes unstable, with intermittent excursions in the lower half of the channel. The velocity profile becomes flatter, more symmetric and $\tilde{u}$ still decreases ($\tilde{u}=0.82$). Finally, for $A = 0.276$ (first row), the flow becomes turbulent with large mixing over the entire cross section, leading to a typical turbulent plug profile ($\tilde{u}=0.8$).

Thanks to this analysis, it is now possible to define a quantitative criterion for the onset of turbulence. In the following, we will consider that the flow becomes turbulent when $\tilde{u} \approx 0.8$. For $0.8\lesssim\tilde{u}<1$, the flow is in an intermediate state dominated by hairpin vortices.

\begin{figure}[htb]
\includegraphics[width=\columnwidth]{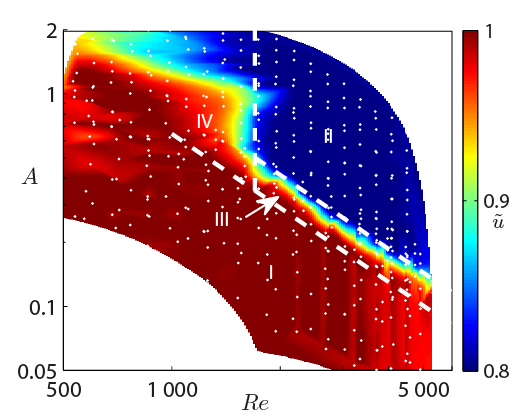}
\caption{Contour plot in log-log scale of $\tilde{u}(Re,A)$ showing the power law for the onset of turbulence for $Re>1500$. \textbf{I}: laminar state, \textbf{II}: turbulent state, \textbf{III}: Regime dominated by hairpin vortices and \textbf{IV}: Below $Re=1500$ it is not possible to sustain turbulence by this mean.}
\label{fig:SurfaceLog}
\end{figure}

We plot on figure \ref{fig:SurfaceLog} the state diagram of the flow, $\tilde{u}(Re,A)$, as a contour plot in a log-log representation. We define four different regions representing different states of the flow. In region \textbf{I}, $\tilde{u}=1$, the flow is steady and the mean velocity profile is parabolic. This region corresponds to the laminar state. Region \textbf{II} ($\tilde{u}\approx0.8$) corresponds to the turbulent state, with high mixing and a plug velocity profile. Region \textbf{III} ($0.8\lesssim\tilde{u}<1$) corresponds to a narrow intermediate state dominated by stable hairpin vortices, similar to those observed in jets in cross flow in a laminar boundary layer  \cite{ilak2010stability}. In region \textbf{IV}, perturbations induced by continuous jets do not trigger the transition but we observe some long-lived flows different from the well known parabolic profile.

\begin{figure}[htb]
\includegraphics[width=\columnwidth]{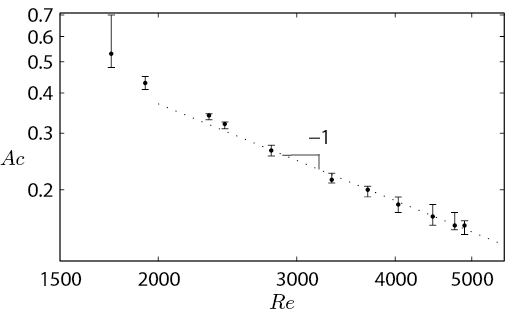}
\caption{\label{fig:Scaling} Log-log plot of the stability curve for the onset of turbulence in plane Poiseuille flow. Squares correspond to $\tilde{u}=0.81$ and solid line is the -1 slope. The minimal amplitude triggering the transition scales like $Re^{-1}$ at higher $Re$ numbers.}
\end{figure}

For $Re>1500$, we define the transition line as the separation between regions \textbf{II} and \textbf{III}, corresponding to $\tilde{u}=0.81$. We show, in figure \ref{fig:Scaling} a log-log plot of the minimal amplitude of the perturbation $A_c = (u_{jet}/u_{cl})_c=f(Re)$ triggering the transition, defined as $\tilde{u}(A_c)=0.81$. The $-1$ slope proposed by Waleffe is added on the plot, showing  good agreement for $Re>2000$. For $Re<2000$, the asymptotic regime is not reached and experimental points deviate from the $-1$ slope. Using $A_c$ for $Re<2000$ to evaluate $\gamma$, would lead to an underestimate of the exponent \cite{cohen_2007}.

The sub-critical transition of plane Poiseuille flow has been studied quantitatively through extensive PIV and LIF measurements. A well-defined state variable, $\tilde{u}$, has been introduced. Thanks to a large number of $\tilde{u}$ measurements, we could define four different regimes of the flow. We then focused on the minimal amplitude of the perturbation, $A_c$, triggering the transition. We found that $A_c(Re)$ scales like $Re^{-1}$ in agreement with the asymptotic non linear theoretical model proposed by Waleffe for shear flows \cite{waleffe2005transition}. We also show for the first time the role played by hairpin-like vortices in each step of the transition.

\begin{acknowledgments}
The authors would like to thank the DGA  for its support, and Dwight Barkley and Laurette Tuckerman for helpful discussions.
\end{acknowledgments}

\bibliography{PRL_GL_JLA_JEW_final}

\end{document}